\documentclass[aps,twocolumn,longbibliography,nofootinbib]{revtex4-1}

\usepackage[colorlinks=true,citecolor=blue,urlcolor=blue,linkcolor=blue,bookmarksopen]{hyperref}
\usepackage{amsmath,amssymb,bm}
\usepackage[dvips]{graphicx}
\usepackage{hyperref}
\usepackage{yfonts}
\usepackage{color}
\newcommand{\beq}{\begin{eqnarray}}
\newcommand{\eeq}{\end{eqnarray}}
\newcommand{\SU}{\text{SU}}
\newcommand{\U}{\text{U}}
\renewcommand\d{\partial}

\begin{document}

\title{Floquet vacuum engineering: Laser-driven chiral soliton lattice in the QCD vacuum}

\author{Akihiro Yamada and Naoki Yamamoto}
\affiliation{Department of Physics, Keio University, Yokohama 223-8522, Japan}

\begin{abstract}
What happens to the QCD vacuum when a time-periodic circularly polarized 
laser field with a sufficiently large intensity and frequency is applied?
Based on the Floquet formalism for periodically driven systems
and the systematic low-energy effective theory of QCD, we show that 
for a sufficiently large frequency and above a critical intensity, 
the QCD vacuum is unstable against the chiral soliton lattice of pions; 
a crystalline structure of topological solitons that spontaneously breaks parity 
and continuous translational symmetries. 
Our work would pave the way for novel ``Floquet vacuum engineering."
\end{abstract}
\maketitle

\section{Introduction}
The quantum vacuum has rich structures, such as quark confinement, 
chiral symmetry breaking, and Higgs condensation. 
Exploring its possible phase transitions under extreme conditions, which
could have occurred in the early Universe for example, is one of the 
important questions in high-energy physics. 
The properties of the vacuum may also be changed by experimental 
methods, which T.~D.~Lee called ``vacuum engineering" \cite{Lee}. 
As one such conventional way, relativistic heavy-ion collisions have been 
used to study the phase structure of QCD 
\cite{Stephanov:2004wx, Fukushima:2010bq}.

Recently, there have been vigorous activities in the area of condensed matter 
physics to realize novel quantum states of materials under control using 
a time-periodic laser field. According to Floquet theory, 
the nonequilibrium quantum many-body dynamics of a time-dependent 
Hamiltonian can be described by a time-independent effective Hamiltonian,
where quasiequilibrium states can appear emergently. 
This is the concept of Floquet engineering~\cite{Bukov2015, Oka2019}.

Although the maximum intensity of the laser is still limited experimentally 
at the moment, one can ask, as a matter of principle, 
what happens to the QCD vacuum when a time-periodic circularly polarized 
laser with a sufficiently large intensity and frequency is applied.

In this paper, we address this question by applying the Floquet theory 
to the QCD vacuum.%
\footnote{For previous works on the applications of the Floquet theory to the 
${\cal N}=2$ supersymmetric QCD based on the AdS/CFT correspondence, 
see Refs.~\cite{Hashimoto:2016ize, Kinoshita:2017uch}.}
Based on the systematic low-energy effective theory, we show that the 
QCD vacuum is unstable against the chiral soliton lattice (CSL) of pions; 
a crystalline structure of topological solitons with spontaneous 
breaking of parity and continuous translational symmetries.
A similar CSL is known to appear in various systems ranging from 
condensed matter to high-energy physics, such as 
chiral magnets \cite{Dzyaloshinsky:1964dz, kishine2015}, 
cholesteric liquid crystals \cite{Gennes1968}, 
and dense QCD matter in a strong magnetic field \cite{Brauner:2016pko} 
or global rotation \cite{Huang:2017pqe, Nishimura:2020odq}; 
see also Refs.~\cite{Brauner:2019rjg, Brauner:2019aid} for 
the CSL in QCD-like theories.
This work would provide a new possibility of the ``Floquet vacuum engineering."

In this paper, we use the natural units $\hbar = c = 1$.

\section{Floquet theory and high frequency expansion}
We first review the Floquet formalism applied to many-body systems.
We are interested in the quantum dynamics described by
\beq
\label{Schrodinger}
{\rm i} \d_t |\Psi(t) \rangle = H(t)|\Psi(t) \rangle,
\eeq
where $|\Psi(t) \rangle$ is the quantum state at time $t$ and 
$H(t) = H_0 + H_{\rm laser}(t)$ is the total Hamiltonian, 
with $H_0$ being the time-independent part of the Hamiltonian and 
$H_{\rm laser}(t)$ the laser-induced time-dependent Hamiltonian satisfying 
$H_{\rm laser}(t+T) = H_{\rm laser}(t)$. Here, the periodicity $T$ is related to 
the driving frequency $\Omega$ via $T = 2\pi/\Omega$. 

The Floquet theorem, which is the temporal version of the Bloch theorem
for spatially periodic crystals, states that the solution of the time-periodic 
Hamiltonian above can be expressed as
\beq
|\Psi(t) \rangle = {\rm e}^{- {\rm i} \varepsilon t} |\Phi(t) \rangle, \quad 
|\Phi(t + T) \rangle = |\Phi(t) \rangle,
\eeq
where $\varepsilon$ is the Floquet quasienergy.

As the total Hamiltonian is periodic, $H(t+T) = H(t)$,
we can perform the discrete Fourier transform as 
\beq
H(t) = \sum_{n} {\rm e}^{- {\rm i} n \Omega t} H_{n}, \quad 
|\Phi(t) \rangle =  \sum_{n} {\rm e}^{- {\rm i} n \Omega t} |\Phi_n \rangle,
\eeq
with $n$ an integer, 
and then Eq.~(\ref{Schrodinger}) can be mapped to a time-independent 
eigenvalue problem \cite{shirley1965solution, sambe1973steady}, 
\beq
\sum_m (H_{n-m} - m \Omega \delta_{mn}) |\Phi_m \rangle = \varepsilon |\Phi_n \rangle.
\eeq

This eigenvalue problem may be solved by a perturbative expansion in terms of 
$1/\Omega$ by assuming that the time scale of interest is much larger than $T$.
The resulting static effective Hamiltonian up to the first order in $1/\Omega$ is
\cite{Mananga2011, Goldman:2014xja}
\begin{gather}
H_{\rm eff} = H_{\rm eff}^{(0)} + H_{\rm eff}^{(1)} +O\left(\frac{1}{\Omega^2} \right)\,,
\nonumber \\
\label{H_eff}
H_{\rm eff}^{(0)} = H_0, \quad H_{\rm eff}^{(1)} = \frac{1}{\Omega} \sum_{n>0} \frac{[H_{-n}, H_{n}]}{n} \,,
\end{gather}
where $H_{\rm eff}^{(n)}$ denotes the $n$th order effective Hamiltonian in 
the expansion of $1/\Omega$.

Note that this truncated effective Hamiltonian describes the transient dynamics 
for certain long but finite time scale \cite{kuwahara2016floquet, mori2016rigorous}. 
In this sense, the ``ground state" that we will discuss based on this effective 
Hamiltonian should be regarded as a quasiequilibrium state.

\section{Application to Dirac fermions and Chern-Simons current}
Let us first apply the Floquet formalism above to generic Dirac fermions (with charge $e$) 
coupled to the laser field in three spatial dimensions.%
\footnote{See also Ref.~\cite{Ebihara:2015aca} for the application of the 
Floquet theory to Dirac fermions. Here, we demonstrate the direct relevance 
of the Chern-Simons current in the Floquet effective Hamiltonian for 
a more generic position-dependent gauge field, which is new to the best 
of our knowledge. }
In the following, we adopt the Coulomb gauge ${\bm \nabla} \cdot {\bm A} = 0$. 
The time-dependent coupling to the laser field is given by
\beq
H_{\rm laser}(t) = -e \int {\rm d}^3 {\bm x} \ \psi^{\dag} \gamma^0 {\bm \gamma} \cdot {\bm A} \psi,
\eeq
where $\gamma^{\mu}$ ($\mu = 0,1,2,3$) are the gamma matrices and 
${\bm A} = {\bm A}(t,{\bm x})$ is the time-periodic gauge field, which we take to be
\beq
{\bm A}(t,{\bm x}) = {\bm A}^{+}({\bm x}) {\rm e}^{- {\rm i} \Omega t} + {\bm A}^{-}({\bm x}) {\rm e}^{{\rm i} \Omega t}.
\eeq

Inserting $H_{\pm 1}$ into Eq.~(\ref{H_eff}) and using the identity
\beq
2 {\rm i} \Omega {\bm A}^{+} \times {\bm A}^{-} = {\bm E} \times {\bm A},
\eeq
we have
\begin{align}
\label{H_eff_1}
H_{\rm eff}^{(1)} 
= -\frac{e^2}{\Omega^2} \int {\rm d}^3 {\bm x} ({\bm E} \times {\bm A}) \cdot {\bm j}^5
\,,
\end{align}
where $j^{5 \mu} = \bar \psi \gamma^{\mu} \gamma^5 \psi$ is the axial current.
By using the Chern-Simons current defined by
$j_{\rm CS}^{\mu} = \frac{1}{2} \epsilon^{\mu \nu \alpha \beta} A_{\nu} F_{\alpha \beta}$, 
Eq.~(\ref{H_eff_1}) can be written as the form of the current-current interaction 
\beq
\label{current-current}
H_{\rm eff}^{(1)} = -\frac{e^2}{\Omega^2} \int {\rm d}^3 {\bm x} \ {\bm j}_{\rm CS} \cdot {\bm j}^5\,.
\eeq
Recall that the axial current $j^{5 \mu}$ carries the helicity of fermions 
while the Chern-Simons current $j_{\rm CS}^{\mu}$ carries the helicity of 
electromagnetic fields. Hence, through this interaction term, the helicity 
of the laser induces the helicity of matter sector composed of Dirac fermions.

For definiteness, below we take 
\beq
\label{A(t,x)}
{\bm A}(t, {\bm x}) = \frac{F}{\Omega} (\cos \Omega(t-z), \lambda \sin \Omega(t-z), 0)\,,
\eeq
where $F$ is the amplitude of the electromagnetic fields (see below) 
and $\lambda = \pm 1$ corresponds to the helicity of the gauge field. 
The corresponding ${\bm A}^{\pm}$ read
\beq
{\bm A}^{+}({\bm x}) = \frac{F}{2\Omega} {\rm e}^{{\rm i} \Omega z} {\bm e}_{\pm}\,, 
\quad
{\bm A}^{-}({\bm x}) = \frac{F}{2\Omega} {\rm e}^{-{\rm i} \Omega z} {\bm e}_{\mp}\,, 
\eeq
for $\lambda = \pm 1$, respectively, where ${\bm e}_{\pm} = {\bm e}_x \pm {\rm i} {\bm e}_y$ 
are the helicity basis with ${\bm e}_{x,y,z}$ being the unit vectors in the $x,y,z$ directions. 
Note that the gauge field above satisfies the transversality condition 
${\bm e}_z \cdot {\bm A} = 0$.

In this case, the electromagnetic components of the laser field are
\begin{align}
\label{E}
{\bm E}(t,{\bm x}) &= F (\sin \Omega(t-z), -\lambda \cos \Omega(t-z), 0), \\
\label{B}
{\bm B}(t,{\bm x}) &= F (\lambda \cos \Omega(t-z), \sin \Omega(t-z), 0), 
\end{align}
and the Chern-Simons current is
\beq
{\bm j}_{\rm CS} = \lambda \frac{F^2}{\Omega} {\bm e}_z\,.
\eeq
Then, the effective Hamiltonian density~(\ref{current-current}) becomes
\beq
\label{H_eff_CS}
{\cal H}_{\rm eff}^{(1)} = -\lambda \frac{e^2 F^2}{\Omega^3} j_z^5\,.
\eeq

Alternatively, one might take a position-independent gauge field
\beq
\label{A(t)}
{\bm A}(t) = \frac{F}{\Omega} (\cos \Omega t, \lambda \sin \Omega t, 0)\,,
\eeq
which also leads to the same Chern-Simons current and the effective Hamiltonian 
density as above.
However, the choice of the position-dependent gauge field~(\ref{A(t,x)}), 
which involves not only the rotating electric field but also the rotating magnetic field, 
is essential in order \emph{not} to cause the Schwinger effect \cite{Schwinger:1951nm}, 
as we will discuss later.

\section{Low-energy effective theory of QCD in the laser field}
We now consider the vacuum of two-flavor QCD in the laser field. 
We are interested in the low-energy regime, whose dynamical degrees of freedom 
are pions---the (pseudo) Nambu-Goldstone (NG) modes associated with chiral symmetry breaking. 
The QCD dynamics in this regime is described by the effective field theory for 
pions, called the chiral perturbation theory 
(see, e.g., Ref.~\cite{Scherer:2002tk} for a review).
Since the generally strong and rapidly changing external electromagnetic field 
explicitly breaks the (approximate) coset symmetry 
$[\SU(2)_{\rm R} \times \SU(2)_{\rm L}]/\SU(2)_{\rm V}$
down to $\U(1)_{\rm A}^{I_3}$, we will focus on the light and slow $\pi_0$ sector. 
We will thus need to consider the original time-dependent part of the 
Hamiltonian for $\pi_0$ augmented by the laser-driven static Floquet 
Hamiltonian for $\pi_0$.

Following the general idea of the low-energy effective field theory, 
we perform a systematic expansion in terms of 
$\epsilon_1 \equiv p/\Lambda$ and $\epsilon_2 \equiv m_{\pi}/\Lambda$, 
where $p$ is the typical energy scale of pions, 
$\Lambda = 4\pi f_{\pi}$ is the energy scale for chiral symmetry breaking 
(with $f_{\pi}$ being the pion decay constant),
and $m_{\pi}$ is the pion mass in the QCD vacuum.
As already described above, we also need to perform the expansion in terms 
of $\epsilon_3 \equiv p/\Omega$. 
In addition, one may set up certain counting for the terms involving $eF$. 
As $eF$ appears only through the combination $e^2 F^2/\Omega^3$ 
[the coefficient of Eq.~(\ref{H_eff_CS})] in the Floquet Hamiltonian
at leading order in $1/\Omega$, we introduce
$\epsilon_4 \equiv e^2 F^2/(\Omega^3 \Lambda)$ 
as another small parameter. 
To be specific, we adopt the counting scheme that they are all comparable, 
$\epsilon_{1,2,3,4} \sim \epsilon$, 
and we write down the effective theory up to $O(\epsilon^2)$ in this paper.

We first consider QCD without the coupling to the laser field.
The effective theory up to $O(\epsilon^2)$ is
\beq
{\cal L}_{\rm EFT} = \frac{f_{\pi}^2}{4} \left[ {\rm tr}(\d_{\mu} \Sigma \d^{\mu} \Sigma^{\dag}) + m_{\pi}^2 {\rm tr} (\Sigma + \Sigma^{\dag}) \right]\,,
\eeq
where the phases of $\Sigma \in \SU(2)$ describe the pion fields
and tr$(...)$ denotes the trace over the flavor space.

Let us then consider the coupling of QCD to the laser field.
In this case, the current-current interaction (\ref{current-current}) needs to be 
extended to include the color and flavor degrees of freedom as
\begin{align}
\label{H_Floquet0}
{\cal H}_{\rm Floquet} &= -\frac{e^2}{\Omega^2} \sum_{{\rm f}={\rm u,d}} Q_{\rm f}^2
{\bm j}_{\rm CS} \cdot {\bm j}^{5}_{\rm f}
\nonumber \\
&= -\frac{e^2}{\Omega^2} {\bm j}_{\rm CS} \cdot 
\left(\frac{1}{3}{\bm j}^{5}_3 + \frac{5}{18}{\bm j}^{5} \right)\,.
\end{align}
Here, $(Q_{\rm u}, Q_{\rm d}) = (2/3, -1/3)$ is the quark charge,
$j^{5\mu}_{\rm u} = \bar {\rm u} \gamma^{\mu} \gamma^5 {\rm u}$ and  
$j^{5\mu}_{\rm d} = \bar {\rm d} \gamma^{\mu} \gamma^5 {\rm d}$ are
axial currents for up and down quarks, and
$j^{5\mu}_{a} = \bar q \gamma^{\mu} \gamma^5 \frac{\tau_a}{2} q$ and
$j^{5\mu} = \bar q \gamma^{\mu} \gamma^5 q$ are flavor-triplet and 
singlet axial currents, respectively,  
with $\tau_a$ ($a=1,2,3$) being the $\SU(2)$ generators 
satisfying the normalization ${\rm tr}(\tau_a \tau_b) = 2 \delta_{ab}$.

At low energy, the axial current $j^{5\mu}_{a}$ is carried by 
pions $\pi_a$ and can be expressed in terms of $\Sigma$ as 
\beq
\label{j5}
j^{5\mu}_{a} = -{\rm i} \frac{f_{\pi}^2}{4}{\rm tr}[\tau_a (\Sigma \d^{\mu} \Sigma^{\dag} + \d^{\mu} \Sigma^{\dag} \Sigma)]\,.
\eeq
Since only the $a=3$ component is relevant for the Floquet Hamiltonian (\ref{H_Floquet0}), 
we set $\Sigma = {\rm e}^{{\rm i} \tau_3 \pi_0}$, 
for which $j^{5\mu}_{3} = - f_{\pi}^2 \d^{\mu} \pi_0$.%
\footnote{We ignore the contribution of the flavor-singlet axial current $j^{5\mu}$ 
carried by the heavy $\eta$ meson.}
By requiring the UV-IR matching for the term (\ref{H_Floquet0}), 
the Floquet Hamiltonian density in terms of $\pi_0$ becomes
\beq
\label{H_Floquet}
{\cal H}_{\rm Floquet} = -\lambda \frac{e^2 F^2 f_{\pi}^2}{3\Omega^3} \d_z \pi_0\,,
\eeq
which is also $O(\epsilon^2)$ in our counting scheme.

The Floquet Hamiltonian (\ref{H_Floquet}) leads to a nontrivial consequence 
in the $\pi_0$ sector, on which we focus below. 
The total low-energy effective Hamiltonian density is then given by 
${\cal H}_{\rm total} = {\cal H}_{\pi_0} + {\cal H}_{\rm Floquet}$, 
where
\beq
{\cal H}_{\pi_0} = \frac{f_{\pi}^2}{2} ({\bm \nabla} \pi_0)^2 + m_{\pi}^2 f_{\pi}^2(1-\cos \pi_0)\,.
\eeq
Here, we also included the offset $m_{\pi}^2 f_{\pi}^2$ such that 
$\langle {\cal H}_{\pi_0} \rangle = 0$ for $\langle \pi_0 \rangle = 0$, 
with $\langle ... \rangle$ being the vacuum expectation value.
Note that the Wess-Zumino-Witten term of the form $\pi_0 {\bm E} \cdot {\bm B}$
\cite{Wess:1971yu, Witten:1983tw} is irrelevant here since ${\bm E} \perp {\bm B}$ 
in the present setup.

The total Hamiltonian density is trivially minimized in the $x$ and $y$ directions
by taking $\langle \d_x \pi_0 \rangle = \langle \d_y \pi_0 \rangle = 0$. 
Then, the remaining Hamiltonian density in the $z$ direction is
\beq
\label{H_total}
{\cal H}_{\rm total}^z = \frac{f_{\pi}^2}{2} (\d_z \pi_0)^2 
- \lambda \frac{e^2 F^2 f_{\pi}^2}{3\Omega^3} \d_z \pi_0 
+ m_{\pi}^2 f_{\pi}^2 (1 - \cos \pi_0) \,.
\nonumber \\
\eeq

We observe that the Hamiltonian~(\ref{H_total}) has mathematically the same 
form as those appearing in chiral magnets \cite{Dzyaloshinsky:1964dz, kishine2015}, 
cholesteric liquid crystals \cite{Gennes1968}, and QCD at finite baryon 
chemical potential in a background magnetic field \cite{Brauner:2016pko}
or global rotation \cite{Huang:2017pqe, Nishimura:2020odq};
through the Floquet engineering, this result provides a new universality 
between high-energy physics and condensed-matter physics.
In particular, the Floquet effective Hamiltonian (\ref{H_Floquet}) for pions 
has the same form as the Dzyaloshinskii-Moriya interaction for magnons 
in chiral magnets \cite{Dzyaloshinsky:1964dz, kishine2015} 
and the Wess-Zumino-Witten term for pions in QCD matter at 
finite density \cite{Son:2004tq, Son:2007ny}.
By utilizing the results there, we can also find the analytic solutions for 
the ground state of our Floquet-QCD system. Below we consider 
$\lambda = 1$ as an example.

\section{Chiral helix and helicity transfer}
Let us start with QCD in the chiral limit where $m_{\pi}=0$. In this case, 
the Hamiltonian density (\ref{H_total}) is minimized when
\beq
\label{pi0}
\langle \pi_0 \rangle = \frac{e^2 F^2}{3\Omega^3} z\,,
\eeq
at which the minimum is given by
\beq
\label{H_minimum}
\langle {\cal H}_{\rm total}^z \rangle = -\frac{1}{2} \left(\frac{e^2 F^2 f_{\pi}}{3\Omega^3} \right)^2 < 0\,.
\eeq
This shows that the QCD vacuum is unstable for an infinitesimally small 
amplitude $F$ against the formation of the helical structure of $\pi_0$, 
given by Eq.~(\ref{pi0}), which spontaneously breaks the parity symmetry.
This has the natural interpretation that the QCD vacuum is turned into the 
helical ground state by the helicity of the laser field.

In fact, in this case we have the following helicity transfer relation 
\beq
\label{helicity}
\frac{1}{f_{\pi}^2} {\bm j}^5_3 = \frac{e^2}{3 \Omega^2}{\bm j}_{\rm CS}\,.
\eeq
Note that in usual relativistic quantum field theories, helicity transfer from 
the gauge sector to the matter sector occurs through the effects related 
to the chiral anomaly \cite{Yamamoto:2015gzz}. In the present case, however, 
this new type of helicity transfer can occur owing to the Floquet effective 
interaction (\ref{current-current}).

\section{Chiral soliton lattice and critical intensity}
In the case of QCD with finite quark masses, the equation of motion 
for the Hamiltonian density (\ref{H_total}) is given by
\beq
\d_z^2 \pi_0 = m_{\pi}^2 \sin \pi_0.
\eeq
This is the same equation of motion for a single pendulum and can be analytically 
solved using the Jacobi elliptic function 
(see also Refs.~\cite{kishine2015, Brauner:2016pko}),
\beq
\cos \frac{\pi_0 (\bar z)}{2} = {\rm sn} (\bar z, k)\,,
\eeq 
where $\bar z \equiv m_{\pi} z/k$ is a dimensionless coordinate 
and $k$ ($0 \leq k \leq 1$) is the elliptic modulus.

This solution shows that $\pi_0(\bar z)$ changes from $2(n-1)\pi$ to $2n\pi$ 
for $(2n-1)K(k) \leq \bar z \leq (2n+1)K(k)$ with $n$ being an integer
and $K(k)$ the complete elliptic integral of the first kind.
Hence, the period of the crystalline structure is 
\beq
\ell = \frac{2k K(k)}{m_{\pi}}\,.
\eeq

We can also show that each lattice carries the axial current as a topological charge. 
In fact, we have
\beq
\int_{(2n-1)K}^{(2n+1)K} {\rm d} \bar z j_3^{5z} (\bar z) 
= 2 \pi f_{\pi}^2\,,
\eeq
which is topologically quantized independently of the detailed profile 
of $\pi_0(z)$.

In summary, this ground-state solution breaks parity and continuous translational 
symmetries and has the topological charge; thus, this is the CSL phase similar to 
those in other systems \cite{Dzyaloshinsky:1964dz, kishine2015, Gennes1968, 
Brauner:2016pko, Huang:2017pqe, Nishimura:2020odq} (see also 
Ref.~\cite{sato2016laser} for the laser-driven CSL in multiferroic magnets). 
It is interesting to note that although the microscopic Hamiltonians are quite 
different between QCD and multiferroics, the same type of the ground state 
is naturally realized in the presence of the circularly polarized laser.

Let us now derive the critical intensity for the realization of CSL in the QCD vacuum.
The energy of each lattice per unit area in the $xy$ plane is given by
\begin{align}
\frac{{\cal E}_{\rm lattice}}{S} &= 4 m_{\pi} f_{\pi}^2 \left[ \frac{2 E(k)}{k} + \left( k - \frac{1}{k} \right)K(k) \right] 
- \frac{2\pi e^2 F^2 f_{\pi}^2}{3 \Omega^3}
\nonumber \\
& \equiv G(k)\,,
\end{align}
where $E(k)$ is the complete elliptic integral of the second kind.
The total energy at length $L$ in the $z$ direction is then%
\footnote{Generally, the quasienergy in Floquet systems is defined modulo
multiples of $\Omega$, and possible Floquet-Umklapp scattering processes 
may violate the energy conservation. In the regime we consider, however, 
the energy transfer of the physical degrees of freedom is much smaller 
than $\Omega$, and such Floquet-Umklapp processes do not occur. 
Hence, the total energy here is well defined.}
\beq
{\cal E}_{\rm total} = \frac{V}{\ell} G(k)\,,
\eeq
where $V = LS$ is the volume of the system.

Minimizing ${\cal E}_{\rm total}$ at fixed $V$ with respect to $k$ yields
\beq
\frac{E(k)}{k} = \frac{\pi e^2 F^2}{12 m_{\pi} \Omega^3}\,.
\eeq
From the inequality $E(k)/k \geq 1$ for $0 \leq k \leq 1$, we find the critical
amplitude for the CSL solution,
\beq
F \geq F_{\rm CSL} \equiv \sqrt{ \frac{12 m_{\pi} \Omega^3}{\pi e^2} }\,,
\eeq
and the corresponding critical intensity $I_{\rm CSL} = F_{\rm CSL}^2$.
This is our main result.
In the chiral limit, it reduces to $F_{\rm CSL} = 0$, as is consistent with
the result of the chiral helix above.
For $F = F_{\rm CSL}$, the parameter $\epsilon_4$ satisfies
\beq
\label{consistency}
\epsilon_4 = \frac{12 m_{\pi}}{\pi \Lambda} \ll 1\,,
\eeq
which confirms the consistency of our analysis.
 
One might think that the vacuum may also be potentially unstable against the pair 
production of charged pions $\pi_{\pm}$ due to the Schwinger effect \cite{Schwinger:1951nm} 
in the confined phase.
For the specific choice of the gauge field~(\ref{A(t,x)}), however, one can argue 
on physical grounds that the Schwinger effect does not occur.
For $\pi_{\pm}$ (with the charge $q = \pm e$) initially at rest, the classical trajectories 
after turning on this laser field are given by \cite{meyer2001}
\begin{align}
x(t) &= - \frac{q F}{m_{\pi} \Omega^2} \sin \left(\frac{\Omega t}{\gamma} \right)\,,
\nonumber \\
y(t) &= \frac{q F}{m_{\pi} \Omega^2} \cos \left(\frac{\Omega t}{\gamma} \right)\,,
\nonumber \\
z(t) &= \frac{(qF)^2 t}{(qF)^2 + 2 (m_{\pi} \Omega)^2}\,,
\end{align}
where
\beq
\gamma = 1 + \frac{1}{2} \left(\frac{q F}{m_{\pi} \Omega} \right)^2\,.
\eeq
Hence, in this case the laser field does not tear the pair of $\pi_{\pm}$ apart, 
as a static magnetic field does not \cite{Dunne:2004nc}.
This should be contrasted with the position-independent gauge field (\ref{A(t)}),
for which the rotating electric field can break the pair of $\pi_{\pm}$ apart, 
triggering the Schwinger effect \cite{Kinoshita:2017uch, Takayoshi:2020afs}.

\section{Application to Weyl/Dirac semimetals}
The argument and results so far should also be applied to Weyl/Dirac 
semimetals with chiral symmetry breaking (see, e.g., Ref.~\cite{Wang:2012bgb}).
In such a system, there is a gapless NG mode associated with $\U(1)$ 
chiral symmetry breaking, which may be regarded as an axion field. 
By repeating the same argument as above, we can construct a low-energy effective 
theory for $\theta$ under the laser field (\ref{A(t,x)}) well below the energy gap as
\beq
\label{H_phi}
{\cal H}_{\theta} = \frac{f^2}{2}({\bm \nabla} \theta)^2 
- \lambda \frac{e^2 F^2 f^2}{\Omega^3} \d_z \theta\,.
\eeq
Here, $f$ is some parameter which cannot be fixed by the symmetry argument
alone (similar to $f_{\pi}$ in the chiral perturbation theory).
We can then show that the original insulating phase is unstable against the 
formation of the chiral helix of $\theta$, 
\beq
\langle \theta \rangle = \lambda \frac{e^2 F^2}{\Omega^3}z
\eeq
for $F>0$.

The electromagnetic properties of such a chiral helix phase was previously studied 
in a different context in Ref.~\cite{Yamamoto:2015maz}. If we further turn on 
an external electric field ${\bm E}_{\rm ex}$, it induces the anomalous Hall effect,
\beq
{\bm j}_{\rm AHE} = \frac{e^2}{4\pi^2} \langle {\bm \nabla} \theta \rangle \times {\bm E}_{\rm ex}
= \lambda \frac{e^4 F^2}{4\pi^2 \Omega^3} {\bm e}_z \times {\bm E}_{\rm ex}\,.
\eeq
Also, the dispersion relation of photons in this phase is modified: one of the helicity 
states acquires an energy gap $\Delta$ while the other has a quadratic dispersion,
the latter of which can be regarded as the so-called type-B NG mode of the 
one-form symmetry \cite{Yamamoto:2015maz, Hidaka:2020ucc}.
As a consequence, the chiral helix behaves as a polarizer which can transmit 
electromagnetic waves with only one type of helicity below the gap $\Delta$
(see also Ref.~\cite{Ozaki:2016vwu}).
These predictions may in principle be tested in tabletop experiments.

\section{Conclusion}

In this paper, we have shown that there exists a previously unknown region of 
laser intensity characterized by the pion mass, where the QCD vacuum is altered 
to a novel state of matter, the pion chiral soliton lattice, by the new mechanism. 
For $\Omega \sim \Lambda \sim 1 \ {\rm GeV}$ (corresponding to a yoctosecond laser) 
with $m_{\pi} \sim 0.1 \ {\rm GeV}$, 
we obtain $F_{\rm CSL} \sim (1 \ {\rm GeV})^2$.%
\footnote{Note that this is still within the applicability of the effective theory in that
the parameter $\epsilon_4$ in Eq.~(\ref{consistency}) is sufficiently small.}
This is above the critical intensity of the Schwinger mechanism, characterized 
by the electron mass, where the QED vacuum becomes unstable due to 
electron-positron pair production. 
Therefore, the critical laser intensity of the Schwinger mechanism, which has not 
been experimentally realized so far, is not the ultimate regime of the quantum 
vacuum physics, and it would be interesting to further aim for a laser intensity 
beyond it in the future.

\acknowledgments
We thank Koichi Hattori for discussions on the Schwinger effect.
This work was supported by the Keio Institute of Pure and Applied Sciences 
(KiPAS) project at Keio University and JSPS KAKENHI Grant No.~19K03852.

\bibliography{Floquet_ref}
  
\end{document}